\documentclass[reprint,aps,prd,showpacs,eqsecnum,twocolumn,superscriptaddress,nofootinbib]{revtex4-1}
\usepackage{amsmath,amssymb,graphicx,color,ulem,multirow, inputenc}
\usepackage{hyperref,ulem,url}
\hypersetup{colorlinks=true}

\newcommand{\vx}{\mathbf{x}} 
\newcommand{\vb}{\mathbf{b}} 
\newcommand{\vp}{\mathbf{p}} 

\begin{document}

\title{Implementation of multi-grid Poisson solver in numerical relativity and its application to gravitational collapse of massive star
} 

\author{Kenta Kiuchi}
\affiliation{Max Planck Institute for Gravitational Physics (Albert Einstein Institute), Am M\"{u}hlenberg, Potsdam-Golm, 14476, Germany}
\affiliation{Center for Gravitational Physics and Quantum Information, Yukawa Institute for Theoretical Physics, Kyoto University, Kyoto 606-8502, Japan}

\author{Hirotada Okawa}
\affiliation{Faculty of Software and Information Technology, Aomori University, Tokyo 134-0087, Japan}

\date{\today}

\begin{abstract}
We develop a new grid-based multi-grid Poisson solver in numerical relativity. We report the performance of the multi-grid Poisson solver in the initial value problems for two-puncture black holes, a static spherical neutron star, a uniformly rotating neutron star in equilibrium, and a gravitationally collapsing massive star. As a demonstration, we conduct a numerical-relativity neutrino-radiation-transfer hydrodynamics simulation of the gravitational collapse of the $9M_\odot$ massive star in Ref.~\cite{Aguilera-Dena:2020mfh} up to the core bounce. During the simulation, we employ the constraint-preserving regird prescription with the newly developed multi-grid Poisson solver to improve the resolution. It shows that the baryonic mass, the  Arnowit-Deser-Misner (ADM) mass, and the ADM-like angular momentum are, respectively, preserved with $O(10^{-3})\%$ and $O(10^{-2})$--$O(10^{-1})$\% accuracy. 
\end{abstract}

\maketitle

\section{Introduction}
Numerical Relativity is a unique tool for building a theoretical model of high-energy relativistic astrophysical transients. Particularly, after the first detection of the gravitational waves from the binary black hole merger GW150914~\cite{LIGOScientific:2016aoc} and subsequent multimessenger event from the binary neutron star merger GW170817~\cite{LIGOScientific:2017vwq,LIGOScientific:2017ync,LIGOScientific:2018hze,LIGOScientific:2018cki}, the importance of numerical relativity is enhanced significantly because all the fundamental interactions equally play a role in such an event, and any analytical techniques break down. 

A couple of the cutting-edge numerical relativity simulations are conducted to date for the binary neutron star mergers~\cite{Kiuchi:2022nin,Kiuchi:2023obe,Hayashi:2024jwt,Fujibayashi:2022ftg,Fujibayashi:2020qda,Fujibayashi:2020dvr,Fujibayashi:2020jfr,Fujibayashi:2017puw,Shibata:2021bbj,Shibata:2021xmo,Gieg:2026beb,Daszuta:2026szb,Most:2023sft}, for the black-hole neutron star binary mergers~\cite{Hayashi:2021oxy,Hayashi:2022cdq,Markin:2026eyc,Foucart:2018lhe}, for the binary black hole mergers~\cite{Boyle:2019kee,Scheel:2025jct}, for the gravitational collapse of the massive stars~\cite{Fujibayashi:2022xsm,Fujibayashi:2023oyt,Shibata:2023tho,Shibata:2025gix,Halevi:2025mga,Kuroda:2024xbe,Kuroda:2023mzi}, for the tidal disruption of the white dwarf by the intermediate mass black hole~\cite{Lam:2022yeg}, and for the accretion-induced collapse of the white dwarf~\cite{Kuroda:2025iyj,Cheong:2024hrd,Combi:2025yvs}.

To handle the large length-scale separation from the neutron star or stellar-mass black hole scale to the gravitational wavelength or the entire massive stellar scale, the implementation of a so-called mesh refinement in numerical relativity codes is mandatory. Since the Einstein equation  contains the Hamiltonian and Momentum constraints, the prolongation (restriction), i.e., the interpolation from a coarser (finer) mesh refinement domain to a finer (coarser) mesh refinement domain, is non-trivial in grid-based free-evolution numerical relativity codes~\footnote{This is not the case for the spectral evolution code, such as {\tt SpEC}~\cite{Foucart:2020xkt} and {\tt SpECTRE}~\cite{Deppe:2024ckt}, because, with a basis function and spectral coefficients, one can interpolate a constraint-preserving metric data in arbitrary spatial point. This technique is also used in the initial data solver exporter in numerical relativity, e.g., {\tt FUKA}~\cite{Grandclement:2009ju,Papenfort:2011}.}. If one employs the unconstrained interpolation of the metric, the constraint violation is unavoidable. 

A similar problem is well known in classical magnetohydrodynamic simulations with mesh refinement~\cite{Balsara:2001}. Employing the unconstrained prolongation, such as a polynomial interpolation, for each magnetic field component implies that one {\it independently} specifies the coefficients stemming from the interpolation scheme. However, the divergence-free condition means that there must be a unique relation between the coefficients for the interpolation. 

To see this problem concisely, let's consider a two-dimensional case. We assume that the $x$- and $y$-components of the magnetic field could be prolonged in the four finer cells centered at $(x_j,y_k),(x_{j+1},y_k),(x_j,y_{k+1})$ and $(x_{j+1},y_{k+1})$, which are covered by one parent cell centered at $(x_J,y_K)=(x_j+x_{j+1},y_k+y_{k+1})/2$. 
The distribution of the magnetic field inside the coarser domain may be written as
\begin{align*}
B^x(x,y) &= a^x_0 + a^x_x P_1 (x) +  a^x_y P_1 (y) \\
B^y(x,y) &= a^y_0 + a^y_x P_1 (x) +  a^y_y P_1 (y) 
\end{align*}
where $a^{x/y}_0$ and $a^{x/y}_{x/y}$ 
denote, respectively, the averaged $x(y)$-component of the magnetic field and its $x(y)$- spatial derivatives in the coarser cell. $P_1(x)=x-x_J$. 
We also assume the second-order-accurate expression for simplicity, 
but extending to the three-dimensional case or to higher-order accuracy is straightforward. 
One may employ an interpolation scheme to determine $a^{x/y}_{x/y}$, 
which corresponds to the unconstrained polongation mentioned above. However, the divergence-free condition $\partial_A B^A=0$, where $A=x,y$, means the relation $a^x_x + a^y_y=0$ 
should hold, which is not guaranteed in the unconstrained prolongation~\footnote{One may consider $a^x_x+a^y_y=0$ holds if one employs the second-order accurate centered finite difference for $a^x_x$ and $a^y_y$ as $(B^x_{J+1,K}-B^x_{J-1,K})/(2\Delta x)$ and $(B^y_{J,K-1}-B^y_{J,K-1})/(2\Delta y)$, respectively. However, in practice, one needs to introduce a limiter to estimate $a^x_x$ and $a^y_y$ to suppress the numerical oscillations. Also, this is not guaranteed for the higher-order prolongation case. Therefore, $a^x_x+a^y_y=0$ is not guaranteed practically.}.
Therefore, one must determine the coefficients $a^{x/y}_{x/y}$ 
in such a way as to satisfy the divergence-free constraint. 

In practice, it is more complicated because the magnetic field is defined on the cell surface in the constrained-transport scheme, and we also have to consider magnetic flux conservation, i.e., the consistency between the magnetic flux calculated on the finer cell surfaces and that on the coarser cell surface, during the prolongation. Nonetheless, due to the linearity of the divergence-free condition, it is doable (see Ref.~\cite{Kiuchi:2012qv} for the details). However, in numerical relativity, due to the non-linearity of the Hamiltonian and Momentum constraints, it is unlikely to be able to determine the coefficients satisfying the constraints during the prolongation. 

A box-in-box mesh refinement widely used in the numerical relativity community is well designed to avoid this issue by combining prolongation only at the refinement boundary composed of several grid points with a constraint-violation-propagating prescription, such as Z4c~\cite{Hilditch:2012fp}. Namely, the unconstrained prolongation at the refinement boundary introduces a constraint violation in a small portion of the mesh refinement domain, but the constraint-violation-propagating prescription washes it away. However, generating a new, finer domain entirely by uncontrained interpolation of the metric during the simulation will severely deteriorate the numerical solution due to constraint violations mentioned above. On the other hand, for example, to simulate gravitational collapse of a massive star, generating a new domain called {\it regrid} during the simulation is necessary to reduce the computational cost~\cite{Sekiguchi:2010ja,Kuroda:2015bta}. Therefore, one must employ the constraint-preserving regrid. 

One idea is to solve the Hamiltonian and Momentum constraints after the regrid. For this, one must develop a multi-grid Poisson solver. In this paper, we report the implementation of the multi-grid Poisson solver in numerical relativity. We validate the implementation in several standard initial value problems in numerical relativity. We also demonstrate a gravitational collapse simulation of a massive star employing the constraint-preserving regrid. 

The paper is organized as follows. Section~\ref{sec:implementation} devotes for the the implementation of the multi-grid Poisson solver. In Sec .~\ref {sec:validation}, we validate the implementation by solving the initial value problems for two-puncture black holes, a spherical neutron star in equilibrium, and a uniformly rotating neutron star in equilibrium. Section~\ref{sec:demonstration} is for the gravitational collapse simulation of the massive star with the regrid. Section~\ref{sec:summary} summarizes the paper. Throughout the paper, we use $c=G=1$ unit unless otherwise stated, and standard notation for the metric index.

\section{Multi-Grid Poisson Solver}\label{sec:implementation}
\subsection{Implementation}
We present the multi-grid method in this section~\cite{Brandt1977}. The iterative method acts as a low-pass filter for solving a Poisson-type equation. 
Therefore, the iterative method on a coarser grid acts as a smoother for lower-frequency modes on the finer grid. 
The coarse-grid correction effectively transfers error between the finer grid (denoted by the subscript $h$) and the coarser grid (denoted by the subscript $H$).

The constraints in numerical relativity are Poisson-type differential equations containing nonlinear source terms, which are schematically expressed by
\begin{align}
\Delta \psi = S(\psi), 
\end{align}
where $\Delta$ is the flat-Laplacian operator. Accordingly, the Full Approximation Scheme (FAS) is employed to solve nonlinear problems~\cite{Brown2005,Teunissen2019}. The FAS V-cycle is described as follows.

\medskip
\noindent
\textbf{(1) Initialization on the finer grid} \\
An initial guess is prepared on the finer grid, typically set as $\psi_h = 0$.

\medskip
\noindent
\textbf{(2) Pre-smoothing (high-frequency error reduction)} \\
On the finer grid, a relaxation method is applied to the nonlinear equation to obtain an approximate solution $\tilde{\psi}_h$.
We employ the W4 method as the relaxation method~\cite{Okawa2023,Fujisawa2019,Okawa2023b,Ogata2023,Hirai2020,Suzuki2021}. The W4 method accelerates convergence in solving nonlinear equations compared to the Jacobi method, as the dependence of the nonlinear source term on $\psi$ is explicitly incorporated into the iteration process (see Appendix~\ref{sec:W4} in detail). 

\medskip
\noindent
\textbf{(3) Residual computation} \\
The residual is computed as
\[
\tilde{r}_h := S(\tilde{\psi}_h) - \bigtriangleup_h \tilde{\psi}_h,
\]
where $\Delta_h$ is the discretized Laplacian operator, where we employ the second-order finite difference in cell-centered Cartesian coordinates. We do not impose any spatial symmetry. 

\medskip
\noindent
\textbf{(4) Restriction to the coarser grid} \\
The approximate solution and residual are transferred to the coarser grid:
\[
\tilde{\psi}_H := I_h^H \tilde{\psi}_h, \quad
\tilde{r}_H := I_h^H \tilde{r}_h.
\]
The explicit form of the restriction operator $I_h^H$ is given in Appendix~\ref{sec:W4} .

\medskip
\noindent
\textbf{(5) Coarser-grid correction (low-frequency error treatment)} \\
The coarser-grid equation in the FAS formulation is constructed as
\begin{eqnarray}
 \bigtriangleup_H \psi_H = \tilde{f}_H,
\end{eqnarray}
where
\begin{eqnarray}
 \tilde{f}_H := \bigtriangleup_H \tilde{\psi}_H + \tilde{r}_H.
\end{eqnarray}
This equation is solved iteratively for $\psi_H$ on the coarser grid by the W4 method.

\medskip
\noindent
\textbf{(6) Prolongation and correction} \\
The coarser-grid correction is transferred back to the finer grid, and the solution is updated as
\begin{eqnarray}
 \psi_h \leftarrow \tilde{\psi}_h + I_H^h \left( \psi_H - \tilde{\psi}_H \right).
\end{eqnarray}
The explicit form of the prolongation operator $I_H^h$ is given in Appendix~\ref{sec:W4} .

\medskip
\noindent
\textbf{(7) Post-smoothing (error refinement)} \\
The Jacobi method, as a relaxation method, is applied on the finer grid to improve the solution. Then, we go back to step (2).

\medskip

The multi-grid Poisson solver is fully parallelized by Open-MP and MPI. 

\subsection{Validation in Newtonian gravity}
We validate the implementation in the Newtonian constant-density sphere problem:
\begin{align}
\Delta \phi = 4 \pi G \rho_0, \label{eq:Newtonian}
\end{align}
where $G$ is the gravitational constant, and $\rho_0$ is the constant density. Because this problem has an analytic solution (see Appendix~\ref{appdx:NL}), it serves as a benchmark for implementation. 

We use $G=M=1$ unit where $M$ is the rest mass of the constant-density sphere in this problem. We set the radius of the sphere to be $R_s=2$. The grid set-up for the coarsest domain is $x^i_{(1)}\in [-512:512]$, and grid spacing is $\Delta x_{(1)}=1024/(N-1)$. The nested-grid hierarchy is $\Delta x_{(\rm lv)}=\Delta x_{(\rm lv-1)}/2$ with ${\rm lv}=2,\cdots ,9$. The finest nested-domain size is $x^i_{(9)}\in [-2:2]$ which covers the entire sphere. We vary $N=16,32,64,128,256$, and $512$ to check the convergence. 

\begin{figure}
    \centering
    \includegraphics[scale=0.35]{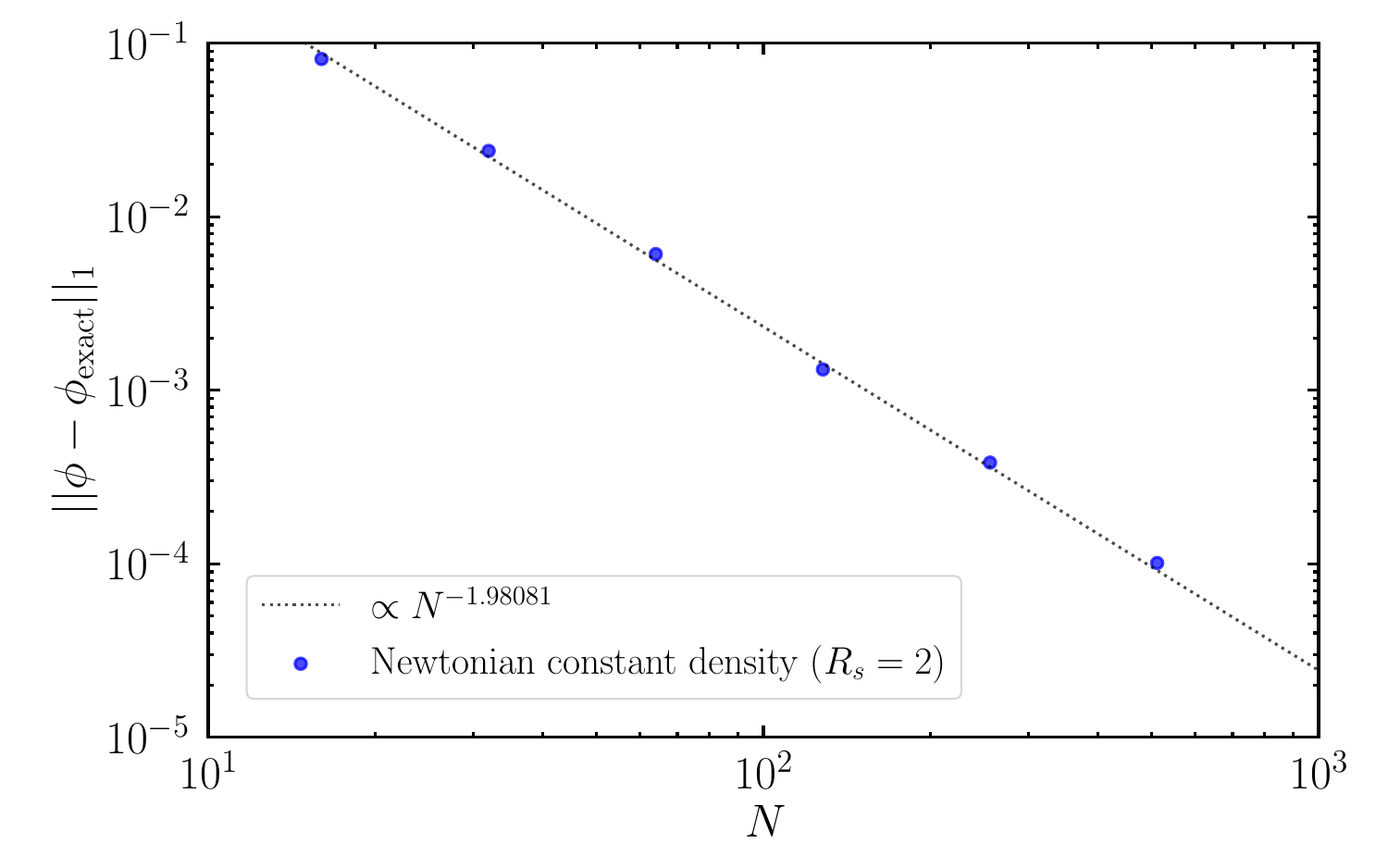}
    \caption{The $L_1$ norm of the error from the exact solution in the Newtonian constant-density sphere problem.}
    \label{fig:MG_vaildation_Newton}
\end{figure}

Figure~\ref{fig:MG_vaildation_Newton} plots the $L_1$ norm for the deviation from the exact solution in this problem. It shows almost second-order convergence, validating the correctness of our implementation. 

\section{Validation in Numerical Relativity}\label{sec:validation}
In this section, we validate the implementation of the multi-grid Poisson solver in numerical relativity. 
First, we solve an initial value problem (IVP) for a two-puncture problem in a vacuum~\cite{Brandt:1997tf}. Second, we solve an IVP for a static spherically symmetric neutron star (Tolman-Oppenheimer-Volkov star)~\cite{1973grav.book.....M}. Third, we solve an IVP for a stationary and uniformly rotating neutron star with the conformal thin-sandwich method~\cite{Paschalidis:2016vmz,York:1998hy,Pfeiffer:2002iy}. 

For the IVP in numerical relativity with the conformal flat assumption, we solve the Hamiltonian constraint:
\begin{align}
\Delta \psi = - \frac{1}{8} \psi^{-7} \bar{A}_{ij} \bar{A}^{ij} - 2\pi \rho_H \psi^5,  \label{eq:Ham}
\end{align}
the Momentum constraint:
\begin{align}
\partial_i {\bar{A}^i}_j - \frac{2}{3} \psi^6 \partial_i K = 8 \pi \psi^6 J_j, \label{eq:Mom}
\end{align}
where $\psi$ denotes the conformal factor. ${\bar{A}^i}_j=\psi^6 {A^i}_j$ where $A_{ij}$ denotes the trace-free part of the extrinsic curvature, and $K$ denotes the trace-part of the extrinsic curvature. $\rho_H=n^\mu n^\nu T_{\mu\nu}$ and $J_i=-{\gamma^\mu}_i n^\nu T_{\mu\nu}$ where $\gamma_{\mu\nu}$, $n^\mu$, and $T_{\mu\nu}$ denote the three spatial metric, the unit normal vector to the hypersurface, and the stress-energy tensor, respectively. 

\subsection{Two puncture}
It is known that Eq.~(\ref{eq:Mom}) under the vacuum condition $\rho_H=J_i=0$ and the maximal slicing condition $K=0$ has an analytic solution describing a black hole spacetime, namely, the puncture solution~\cite{Brandt:1997tf}:
\begin{widetext}
\begin{align}
\bar{A}_{ij} = \sum^2_{I=1} \left[\frac{3}{4r_I^2} \left(p_i^I \bar{n}_j^I + p_j^I \bar{n}_i^I+f^{kl}p_k^I \bar{n}_l^I\left(\bar{n}_i^I\bar{n}_j^I-f_{ij}\right)+\frac{3}{r_I^3}f^{kl}\left(s^I_{ik}\bar{n}^I_l \bar{n}^I_j+s^I_{jk}\bar{n}^I_l \bar{n}^I_i\right)\right)\right],
\end{align}
\end{widetext}
where $I$ denotes a label of the puncture black hole, where we assume two black holes. $f_{ij}$ denotes the spatial component of the Minkowski metric. $r_I=|x^i-x^i_I|$, and $\bar{n}^I_i=(x^i-x^i_I)/r_I$ where $x^i_I$ denotes the puncture position. $p^I_i$ and $s^I_{ij}$ describe the linear momentum and angular momentum, respectively. In this validation problem, we assume $p^{I=1,2}_i=(0,\pm 0.14M,0)$, $s^{I=1,2}_{ij}=0$, and $x^i_{I=1,2}=(0,\pm 3.5M,0)$ where $M$ denotes the total mass of the black holes. Given $\bar{A}_{ij}$, we rewrite Eq.~(\ref{eq:Ham}):
\begin{align*}
&\Delta u = - \frac{1}{8}\psi^{-7} \bar{A}_{ij}\bar{A}^{ij},\\
&\psi = 1 + \frac{M}{4r_1} + \frac{M}{4r_2} + u,
\end{align*}
and solve it with our multi-grid Poisson solver. 

We set the size of the coarsest domain to be $x^i_{(1)} \in [-512M:512M]$, and the grid spacing is $\Delta x_{(1)}=1024M/(N-1)$. The nested-grid hierarchy is $\Delta x_{(\rm lv)}=\Delta x_{(\rm lv-1)}/2$ with ${\rm lv}=2,\cdots,8$. We vary $N=16,32,64,128,256$, and $512$ to check convergence. Figure~\ref{fig:MG_vaildation} plots the $L_1$ norm of the Hamiltonian constraint. We quantify the convergence order by fitting the $L_1$ norm as $\approx 1.56$. The convergence order is slightly worse than the expected value of $2$.  

\subsection{TOV star}
For a TOV star with the maximal slicing condition, the Momentum constraint~(\ref{eq:Mom}) is trivial. Given a matter field, we solve the Hamiltonian constraint~(\ref{eq:Ham}). To model a neutron star, we employ MPA1 as the neutron star matter equation of state~\cite{Muther:1987xaa}, and we assume an Arnowit-Deser-Misner (ADM) mass of the neutron star to be $1.4M_\odot$~\cite{1973grav.book.....M,Gourgoulhon:2007ue}. We utilize the public spectral library FUKA~\cite{Grandclement:2009ju,Papenfort:2011} to export the rest-mass density field of the non-rotating neutron star in equilibrium in the isotropic coordinate with the same mass and the same equation of state. 

The domain size of the coarsest nested grid is $x^i_{(1)}\in[-2048M_\odot:2048M_\odot]$, and we employ the eight nested domains, i.e., ${\rm lv}=1,\cdots ,8$. The grid resolution is $\Delta x_{(1)}=4096 M_\odot/(N-1)$. We employ the 2:1 refinement, and we vary $N=32,64,128,256$, and $512$. Because the coordinate radius of the neutron star is $\approx 6.74 M_\odot$, the finest domain with $x^i_{(8)}\in [-16M_\odot:16M_\odot]$ covers the entire neutron star. 

Left panel of Fig.~\ref{fig:MG_vaildation} plots the $L_1$ norm of the Hamiltonian constraint. The convergence order is almost second order as we expect.  

\subsection{Rapidly rotating neutron star}
Final validation is a uniformly rapidly rotating neutron star in equilibrium. For this problem, we employ the conformal thin-sandwich formulation~\cite{York:1998hy,Pfeiffer:2002iy}. Given the matter field, we solve
\begin{widetext}
\begin{align}
&\Delta \psi = - \frac{1}{8} \psi^5 \tilde{A}^{ij}\tilde{A}_{ij} - 2 \pi \rho_H \psi^5,\\
&\Delta \left(\alpha \psi\right) = 2 \pi \left(\rho_H + 2{S^k}_k \right) \alpha \psi^5 + \frac{7}{8} \tilde{A}^{ij}\tilde{A}_{ij} \alpha \psi^5,\\
&\Delta \beta^i + \frac{1}{3} f^{ij} \partial_j \partial_k \beta^k + 2 \alpha \tilde{A}^{ij} \left(6\partial_j \ln\psi - \partial_j \ln \alpha \right) = 16 \pi \alpha f^{ij} J_j, \label{eq:MD} \\
&\tilde{A}_{ij} = \frac{1}{2\alpha} \left( f_{jk} \partial_i \beta^k + f_{ik} \partial_j \beta^k - \frac{2}{3} f_{ij} \partial_k \beta^k \right)
 \end{align}
\end{widetext}
where we assume the conformal flat, the maximal slicing condition $K=0$ and $\partial_t \tilde{\gamma}_{ij}=0$. $\tilde{A}_{ij}=\psi^{-4}A_{ij}$ and ${S^k}_k=\gamma^{\mu k}{\gamma^\nu}_k T_{\mu\nu}$. 

We employ the Ohara-Shibata decomposition to solve Eq.~(\ref{eq:MD})~\cite{Shibata-textbook}:
\begin{align}
\beta^i = W^i + f^{ij}\partial_j \left(\chi-f_{kl}W^k x^l\right).
\end{align}
With this decomposition, solving Eq.~(\ref{eq:MD}) is equivalent to solve 
\begin{align}
&\Delta \chi = f_{ij} S^i x^j, \label{eq:MD1} \\
&\Delta W^i = S^i, \label{eq:MD2}
\end{align}
where $S^i=16 \pi \alpha f^{ij} J_j- 2 \alpha \tilde{A}^{ij} \left(6\partial_j \ln\psi - \partial_j \ln \alpha \right)$ as long as $S^i$ is differentiable. 

For the matter field, FUKA~\cite{Grandclement:2009ju,Papenfort:2011} exports the rest-mass density and the four velocity for the uniformly rotating star in equilibrium with the ADM mass $=1.4M_\odot$, the ADM-like angular momentum $J_z^\mathrm{ADM-like}=0.392M_\odot^2$~\cite{Gourgoulhon:2007ue} and MPA1 equation of state~\cite{Muther:1987xaa}. The equatorial and polar coordinate radius is $\approx 6.86M_\odot$ and $6.62M_\odot$, respectively. The grid setup is the same as for the TOV star test. 

\begin{figure*}
    \centering
    \includegraphics[scale=0.35]{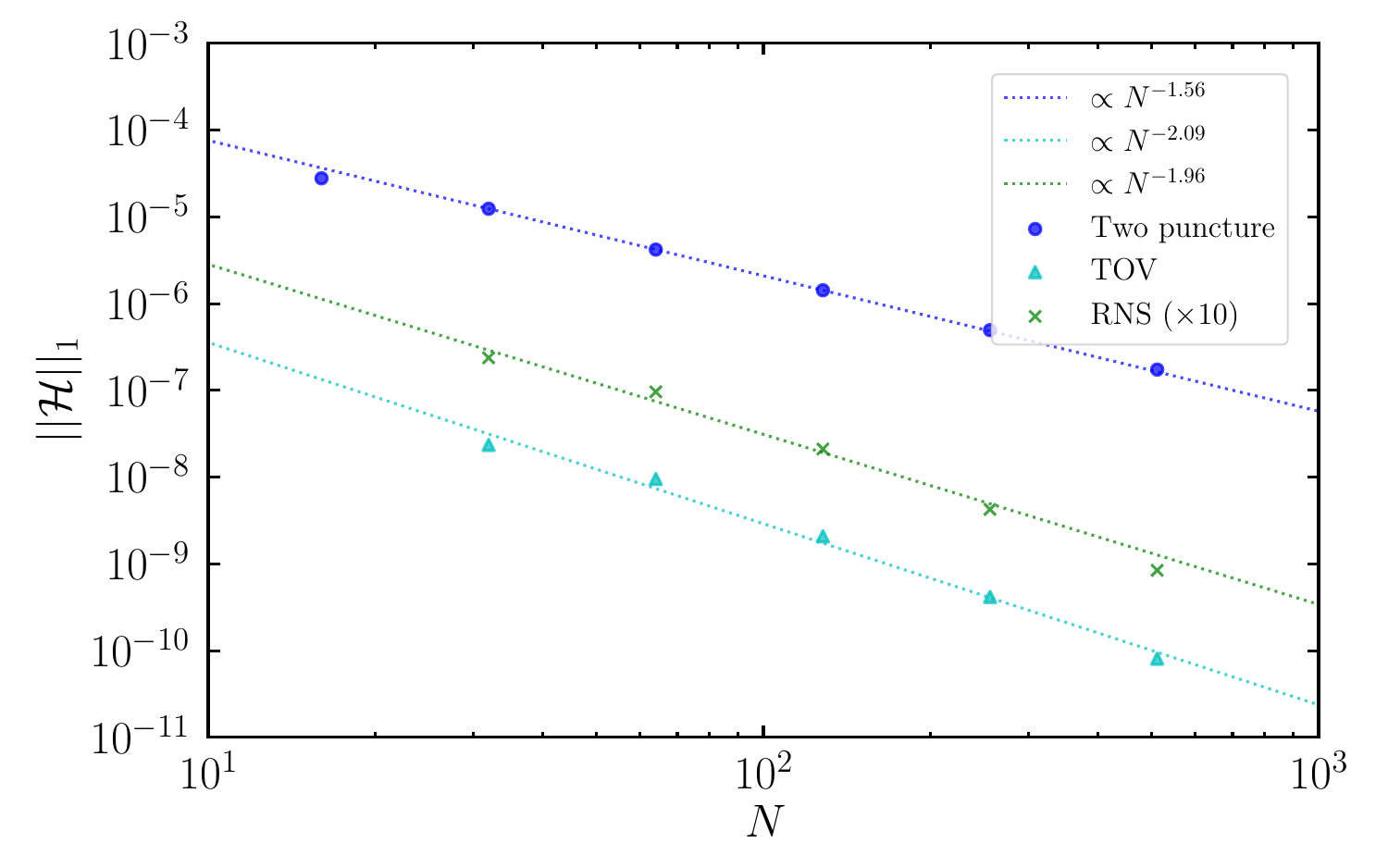}
    \includegraphics[scale=0.35]{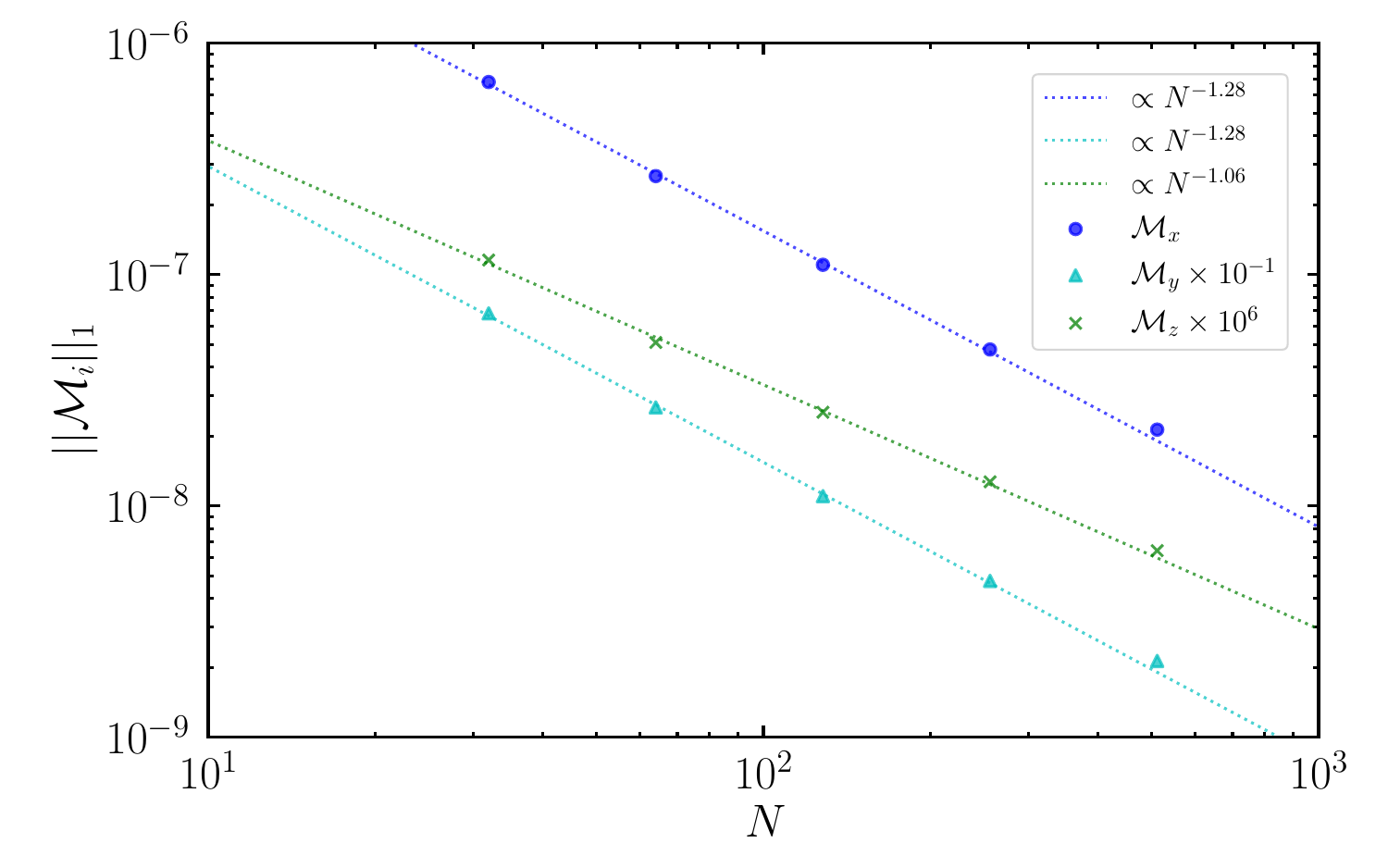}
    \caption{(Left) The $L_1$ norm of the Hamiltonian constraint as a function of the grid number $N$ for the two-puncture black hole (blue), the TOV star (cyan), and the uniformly rotating neutron star (green), respectively. For visibility, the norm is magnified by a factor of $10$ for the uniformly rotating case. 
    (Right) The $L_1$ norm of the Momentum constraint for the uniformly rotating star. The blue, cyan, and green dots represent the $x$, $y$, and $z$ components, respectively. For visibility, the $y$ component is divided by $10$, and the $z$ component is magnified by a factor of $10^6$. }
    \label{fig:MG_vaildation}
\end{figure*}

The left panel of Fig.~\ref{fig:MG_vaildation} plots the $L_1$ norm of the Hamiltonian constraint, and it shows almost second-order convergence. The right panel of Fig.~\ref{fig:MG_vaildation} plots the $L_1$ norm of the Momentum constraint. The $x$ and $y$-components show convergence order of $\approx 1.28$, which is non-negligibly worse than the expected value of $2$. The most likely reason stems from the source $S^i$ in Eqs.~(\ref{eq:MD1})--(\ref{eq:MD2}) could be discontinuous at the stellar surface due to the uniform rotation. 
A multi-domain decomposition or a surface-fitted coordinate could mitigate this problem. However, the implementation of them is beyond the scope of this paper. The $z$-component exhibits a convergence order of $\approx 1.06$. The poor convergence in $z$ components is presumably that it almost reaches machine precision, $\sim 10^{-15}$. 

\section{Application to Core collapse of massive star}\label{sec:demonstration}
In this section, we demonstrate the regrid with the multi-grid Poisson solver for the gravitational collapse simulation of a massive stellar core. 

\subsection{Stellar model and initial condition}
\begin{figure}
    \centering
    \includegraphics[scale=0.35]{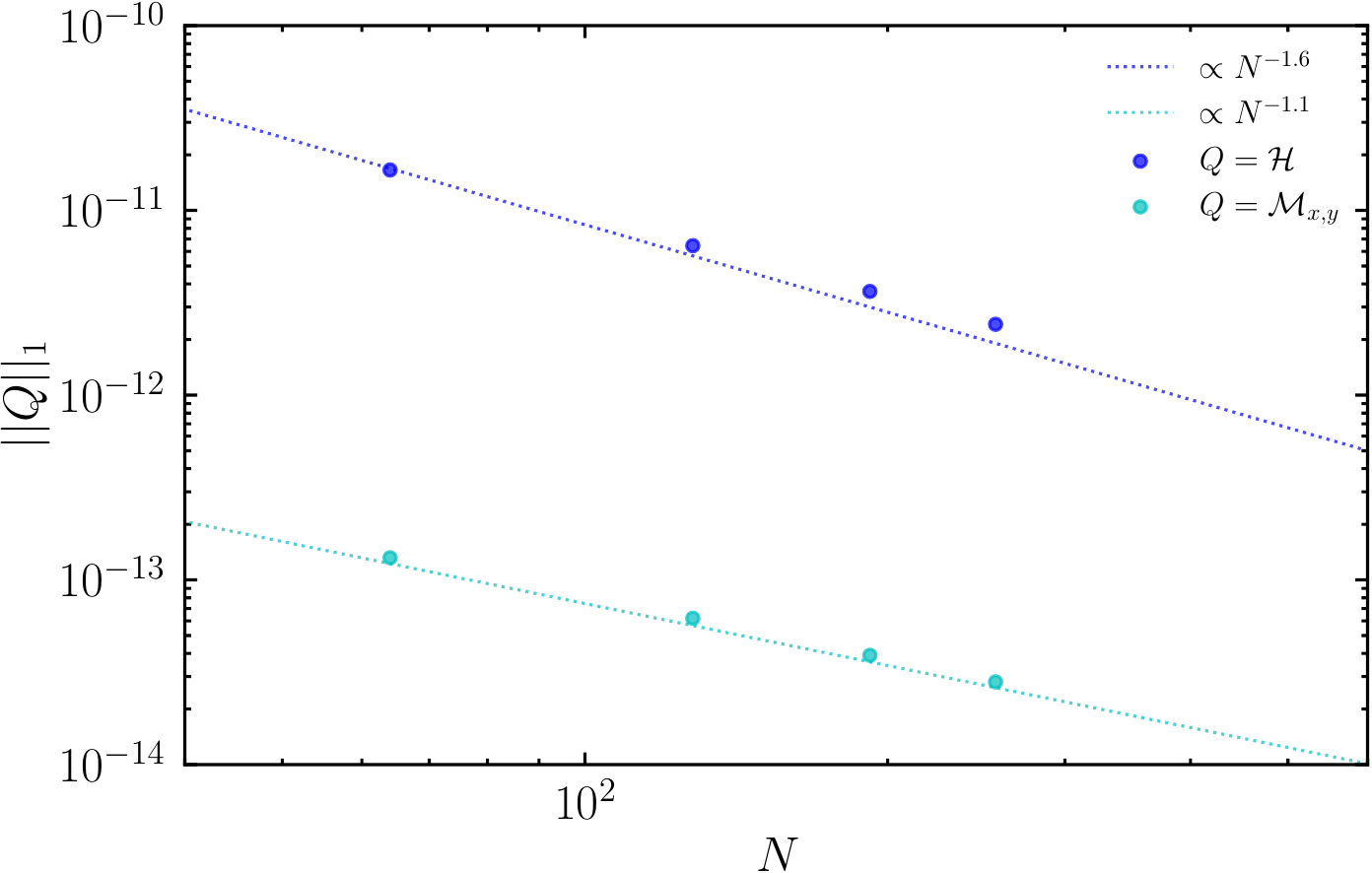}
    \caption{The Hamiltonian and Momentum constraints as a function of the grid number for the initial value problem with the stellar evolution profile {\tt A009}.}
    \label{fig:IVP_SEV}
\end{figure}

We employ the progenitor model of the rotating massive star in Ref.~\cite{Aguilera-Dena:2020mfh}, whose zero-age-main-sequence mass is $9M_\odot$ ({\tt A009} in Ref.~\cite{Fujibayashi:2022xsm}). We export the rest-mass density, velocity, and pressure profiles from the stellar evolution model {\tt A009} to the multi-grid Poisson solver, and solve the IVP with the conformal thin-sandwich formalism. We employ nine nested domains with $\Delta x_{(1)}=2.5\times 10^5/(N-1)~{\rm km}$. We vary $N=256,192,128$, and $64$. Figure~\ref{fig:IVP_SEV} plots the Hamiltonian constraint and the Momentum constraint as a function of $N$. The convergence order is $\approx 1.6$ for the Hamiltonian constraint, and $\approx 1.1$ for the Momentum constraint. Likely in the uniformly rotating neutron star problem, the source term $S^i$ in Eqs.~(\ref{eq:MD1})--(\ref{eq:MD2}), which is discontinuous at the boundary between different chemical composition layers~\cite{Aguilera-Dena:2020mfh}, is presumably a source of the poor convergence order in the Momentum constraint. Nonetheless, the solution converges. 

\subsection{Evolution code and grid setup}
We employ the in-house numerical relativity code {\tt NANASI}~\cite{Kiuchi:2022}. {\tt NANASI} implements the Einstein solver based on Baumgarte-Shapiro-Shibata-Nakamura-puncture formulation~\cite{Shibata:1995,Baumgarte:1998te,Campanelli:2005dd,Baker:2005vv} augmented by the Z4c constraint propagation~\cite{Hilditch:2012fp}. {\tt NANASI} also implements the second-order general relativistic finite-volume HLLC Riemann solver~\cite{Mignone:2005ft,Kiuchi:2022,White:2015omx} augmented by the third-order piecewise parabolic method for the cell reconstruction~\cite{Colella:1984}. 
{\tt NANASI} employs the gray M1+Leakage neutrino radiation transfer, which takes into account neutrino cooling and heating~\cite{Sekiguchi:2010ep,Sekiguchi:2011zd,Sekiguchi:2015dma,Sekiguchi:2016bjd}. A magnetohydrodynamics module is turned off in this paper. As a nuclear equation of state, we employ the DD2 equation of state~\cite{Hempel:2009mc}, which is stitched with the Helmholtz equation of state~\cite{Timmes} to extend its coverage into the low-density, low-temperature region. The minimum density and temperature are $1.67\times 10^{-9}~{\rm g/cm^3}$ and $10^{-7}~{\rm MeV}$, respectively. To cover a wide dynamic range of the massive stellar core collapse problem, {\tt NANASI} implements the 2:1 conservative static mesh refinement based on the Oliger-Berger adaptive mesh refinement algorithm~\cite{Berger:1984zza}. Each nested domain is composed of a concentric Cartesian box with a cell-center grid. {\it Reflux} prescriptions guarantee baryonic mass and the ADM-related quantities conservation with reasonable accuracy (see below). 

We initialize a simulation with nine nested domains with the grid spacing $\Delta x_{(\rm lv)}$ imported from the multi-grid 
Poisson solver. For the dynamical evolution, we impose the octant symmetry to reduce the computational cost. The cell-center grid of the nested domain lv is $x^{(\rm lv)}_j=(j+1/2)\Delta x_{(\rm lv)}$ with $j=0,\cdots,N_d$. To check the convergence, we vary $N_d=121,93$, and $61$. We should note that we export the initial data with $N=256,192,128$ in the multi-grid solver for simulations with $N_d=121,93,61$ in {\tt NANASI}, respectively. 

Once the central maximum density reaches the regrid threshold shown in Table~\ref{tab:regrid_thresh}, we export the {\tt NANASI} simulation data to the multi-grid solver, and increment a finer nested domain. We employ a conservative prolongation for generating the conserved density, momentum, and energy in a new, finer nested domain. It guarantees the baryonic mass conservation, linear momentum conservation, and energy conservation with machine precision. Then, we employ the conformal thin-sandwich method for solving the IVP with this new grid setup, assuming the conformal flat condition and maximal slicing condition, i.e., $\tilde{\gamma}_{ij}=f_{ij}$ and $K=0$, where $\tilde{\gamma}_{ij}$ presents the conformal three metric. After solving the IVP, we export the initial data to {\tt NANASI}, and continue the simulation. We should note that we initialize the conformal three metric and the trace part of the extrinsic curvature as $\tilde{\gamma}_{ij}=f_{ij}$ and $K=0$ every time after the regrid. We repeat this procedure, typically, five times, to achieve the desired resolution in the final finest domain. Table~\ref{tab:regrid_thresh} summarizes our regrid strategy in this demonstration. 

\begin{table*} 
	\centering
	\caption{Threshold rest-mass density for the regrid (second column). Achieved grid resolution with the regrid (fourth-sixth column). The unit in the achieved resolution is $\rm km$. At the beginning of the simulation, the finest grid is $\Delta x_{(9)}$. Once the regrid criterion is satisfied, we increment a new, finer nested domain with grid spacing $\Delta x_{(10)}$, and solve the initial value problem. This procedure is repeated until the total number of the nested domains becomes $14$. 
    } \label{tab:regrid_thresh}
	\begin{tabular}{clccccc} 
		\\
		\hline\hline
        Stage & Regrid criterion & ${\rm lv}_{(\rm max)}$ & $\Delta x_{{\rm lv}_{(\rm max)}}(N=121)$ & $\Delta x_{{\rm lv}_{(\rm max)}}(N=93)$ & $\Delta x_{{\rm lv}_{(\rm max)}}(N=61)$ \\
        \hline\hline
		$1$ & $\rho_{\rm thr}=10^{10}~{\rm g/cm^3}$ & 9  & $3.68$ & $4.98$ & $7.56$ \\
        $2$ & $\rho_{\rm thr}=10^{11}~{\rm g/cm^3}$ & 10 & $1.84$ & $2.49$ & $3.78$ \\
        $3$ & $\rho_{\rm thr}=10^{12}~{\rm g/cm^3}$ & 11 & $0.92$ & $1.25$ & $1.89$ \\
        $4$ & $\rho_{\rm thr}=10^{13}~{\rm g/cm^3}$ & 12 & $0.46$ & $0.62$ & $0.95$ \\
        $5$ & $\rho_{\rm thr}=10^{14}~{\rm g/cm^3}$ & 13 & $0.23$ & $0.31$ & $0.47$ \\
        $6$ & $\rho_{\rm thr}=\infty$               & 14 & $0.12$ & $0.16$ & $0.24$ \\
        \hline\hline
	\end{tabular}
\end{table*}

\subsection{Result}

\begin{figure}
    \centering
    \includegraphics[scale=0.35]{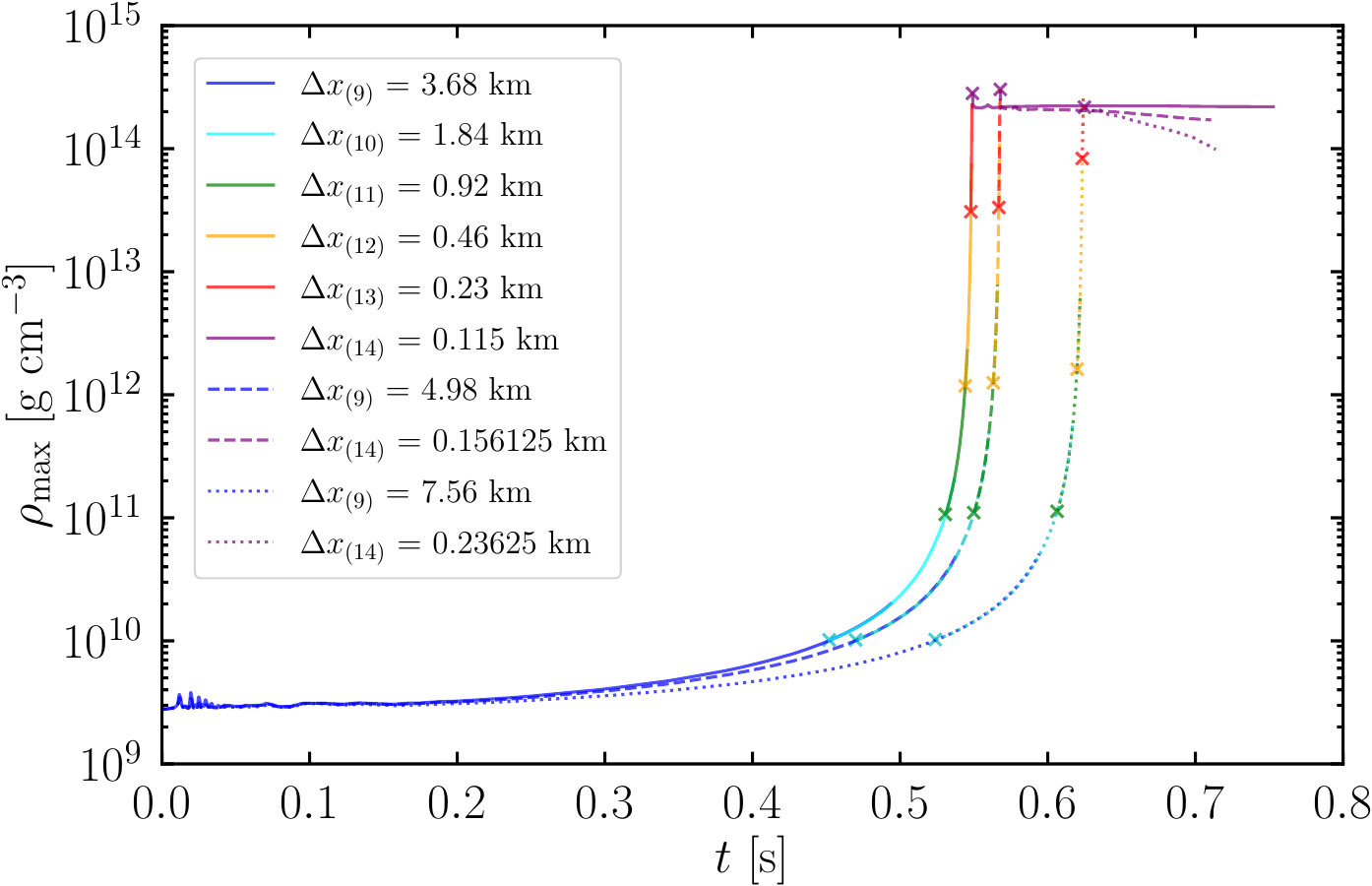}
    \caption{(Left) Central maximum rest-mass density as a function of time. The cyan, green, orange, red, and purple cross symbols present the timing of the regrid.
    The solid, dashed, dotted curves denote $N_d=121,93,$ and $61$, respectively.}
    \label{fig:rhomax}
\end{figure}

Figure~\ref{fig:rhomax} plots the evolution of the central density. To validate the regrid prescription, we continue the simulation with the grid set up before the regrid and compare the result. The cross symbols in the figure present the timing of the regrid. The figure shows that the results with and without regrid overlap nicely (see the overlap after the cross symbols), validating the regrid with our multi-grid solver. Also, the figure shows that the central density of the proto-neutron star remains almost constant after the bounce if we employ the high resolution (see the solid curve). The bounce time is $0.55$~s, $0.57$~s, and $0.64$~s for $N_d=121$, $93$, and $61$, respectively. It converges with the convergence order $\approx 2.3$.

\begin{figure}
    \centering
    \includegraphics[scale=0.35]{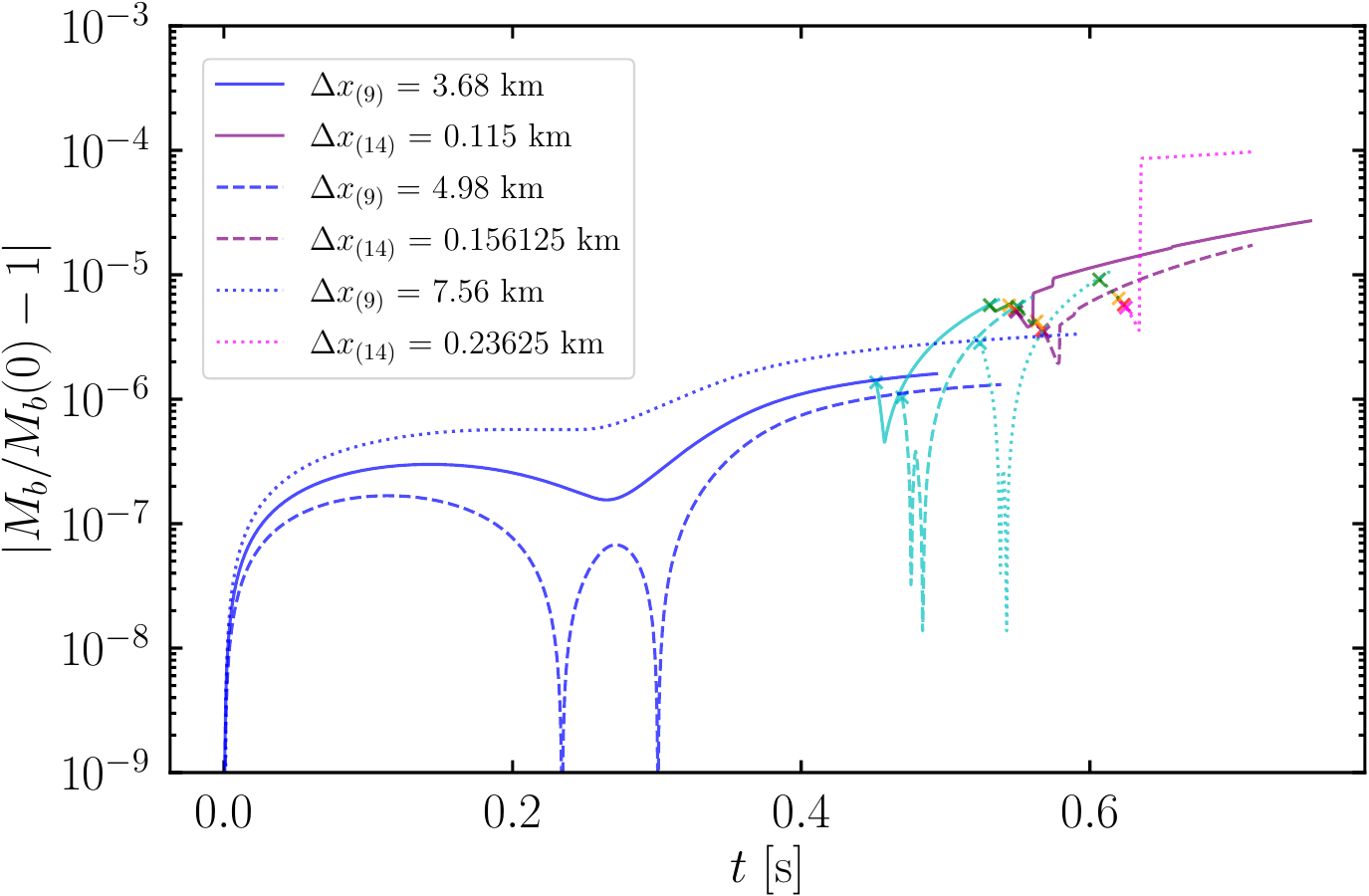}
    \caption{The baryonic mass conservation error as a function of time. The color and style of the curves and cross symbols are the same as Fig.~\ref{fig:rhomax}.}
    \label{fig:Mb}
\end{figure}

Figure~\ref{fig:Mb} plots the violation of the baryonic mass. Although we observe slightly irregular behavior in the baryonic mass conservation error after the regrid, which is presumably related to the atmospheric prescription in the low-density part of the star, it remains $O(10^{-3})$\% at both middle and high resolution after the bounce. 

\begin{figure*}
    \centering
    \includegraphics[scale=0.35]{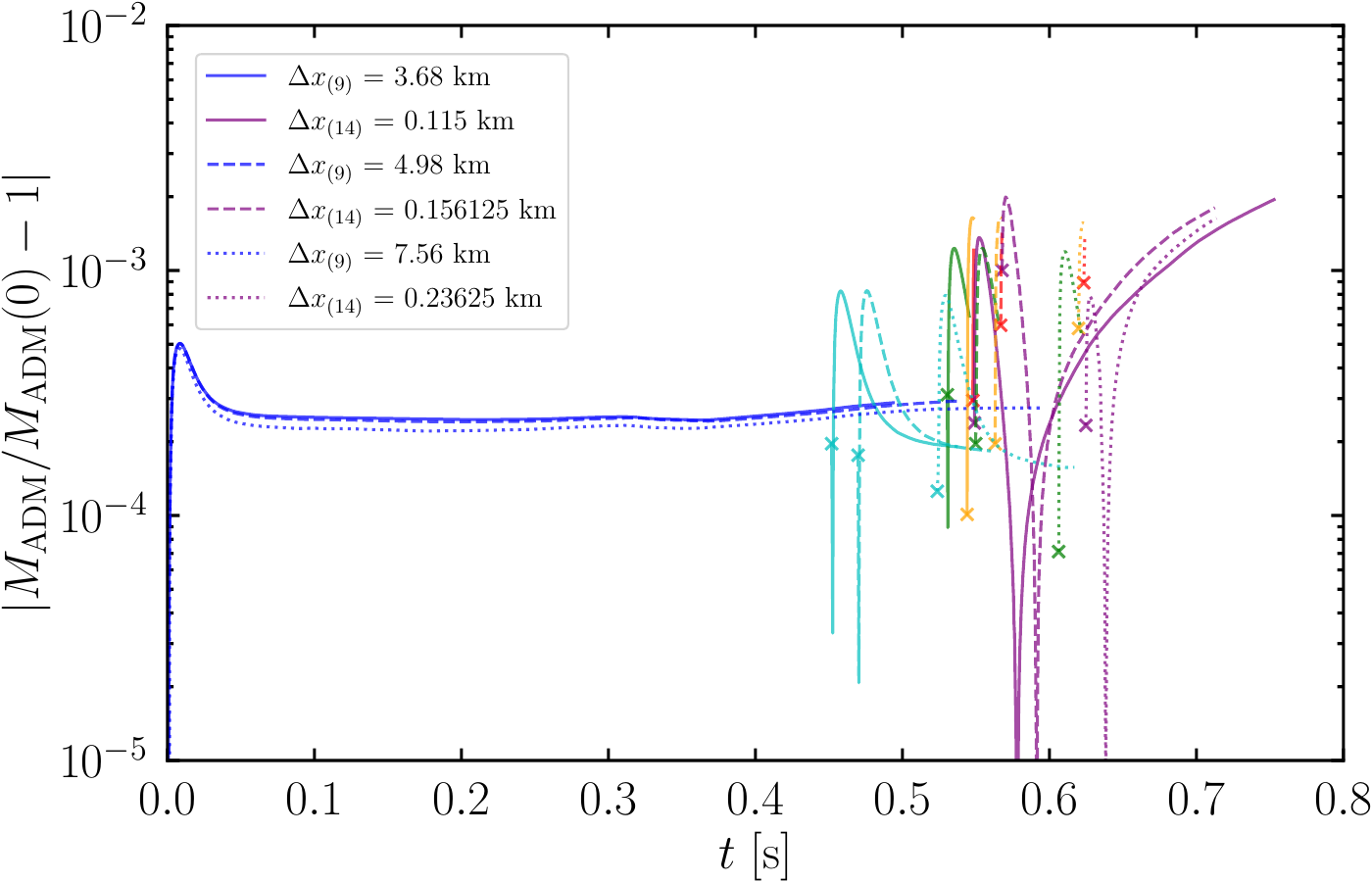}
    \includegraphics[scale=0.35]{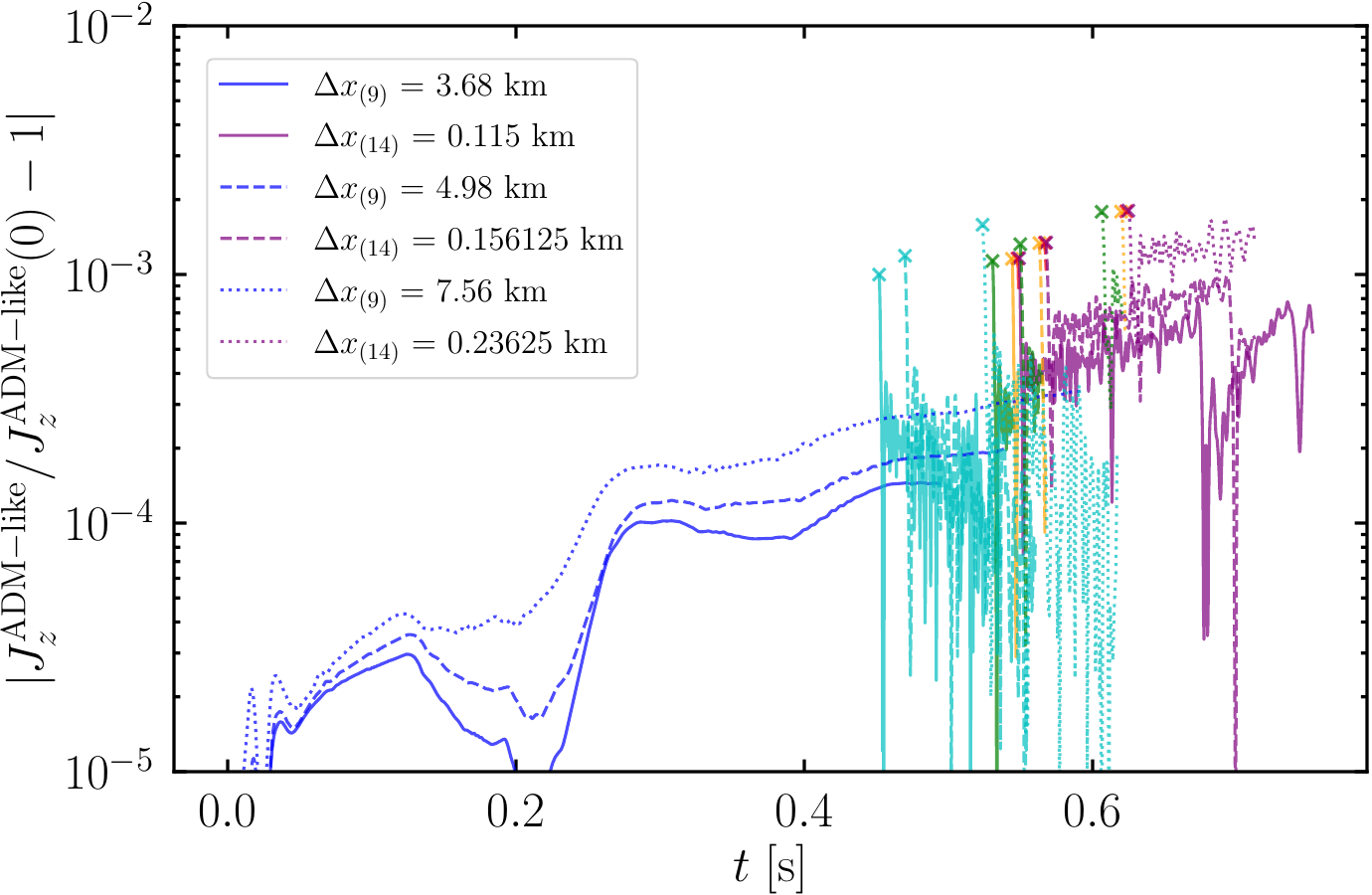}
    \caption{(Left) The ADM mass conservation error as a function of time. (Right) The ADM-like angular momentum conservation error as a function of time. The color and style of the curves and cross symbols are the same as Fig.~\ref{fig:rhomax}.}
    \label{fig:MADM_JADM}
\end{figure*}

Figure~\ref{fig:MADM_JADM} plots the conservation of the ADM mass (left) and ADM-like angular momentum (right). We note that we estimate them by the volume integration rather than the surface integral to improve the accuracy~\cite{Gourgoulhon:2000nn,Shibata-textbook}. We should note that the volume-integrated ADM-related quantities are sensitive to the constraint violations because both constraints are used to rewrite the equation from the surface integral to the volume integral (see Ref.~\cite{Gourgoulhon:2000nn} for example). 
Also, we neglect the contribution from the gravitational wave emission since it should be minor. 
Although the conservation of both quantities is more challenging than the baryonic mass, the violation of $O(10^{-2})$--$(10^{-1})$\% is likely to be acceptable. Also, the violation is improved by improving the resolution. 

We conclude that {\tt NANASI} demonstrates the massive stellar collapse simulation with the regrid prescription with reasonable accuracy.

\section{Conclusion}\label{sec:summary}
We report the implementation of a second-order-accurate multigrid solver in numerical relativity. We validate the code's performance for the two-puncture problem, the TOV star, and the uniformly rotating neutron star in equilibrium, respectively, by solving a couple of non-linear Poisson-type equations in the IVP of numerical relativity. 
We also demonstrate the constraint-preserving regrid in the gravitational collapse simulation of the massive star. 

We plan to simulate the gravitational collapse of a massive star for launching a relativistic jet via a large-scale dynamo.

\acknowledgments 

KK thanks the Computational Relativistic Astrophysics division members in AEI, especially Masaru Shibata. KK also appreciates Yuichiro Sekiguchi and Sho Fujibayashi for a fruitful discussion. Numerical computations were performed on the clusters Sakura, Momiji, Raven, and Viper at the Max Planck Computing and Data Facility, as well as on FUGAKU in RCCS (hp240532, hp250570, hp250066, hp260063). KK is supported by Grant-in-Aid for Scientific Research (grant No.~23K25869) of Japanese MEXT/JSPS, and HO is in part supported by JP23K03222.

\section*{Data availability} 
The data supporting the findings of this article are available upon request.

\section*{Code availability} 
The code is available in the public domain~\cite{Okawa2026}.


\appendix

\section{Multi-grid method in the public code}\label{sec:W4}
We implement the W4 and Jacobi methods as internal relaxation methods in the multi-grid scheme. The W4 method was developed for solving nonlinear systems of equations and has been applied to several problems in astrophysics~\cite{Okawa2023,Okawa2023b,Fujisawa2019,Ogata2023,Hirai2020,Suzuki2021}. In this appendix, we briefly summarize the differences between the W4 method and the Jacobi method for solving nonlinear equations.

\subsection{Jacobi method}

The Jacobi method is a classical iterative scheme for solving discretized equations. 
For a nonlinear equation written as
\begin{eqnarray}
 A\vx = \vb(\vx),\label{eq:av-b}
\end{eqnarray}
where $A \in \mathbb{R}^{N \times N}$ and $b \in \mathbb{R}^N$ denote the discretized matrix of the Poisson equation and the source vector, respectively. The Jacobi iteration updates the solution as
\begin{eqnarray}
 \vx^{(k+1)} = \vx^{(k)} + \omega D^{-1} \left( \vb(\vx^{(k)}) - A\vx^{(k)} \right),
\end{eqnarray}
where $D$ represents the diagonal part of the matrix $A$, and $\omega$ is a relaxation parameter. The Jacobi method only uses local information. 
Therefore, it does not take into account the variation of the solution vector during the iteration. As a result, its convergence rate can be slow, especially when the dependence of the source $\vb(\vx)$ on $\vx$ is strong.

\subsection{W4 method}

In contrast, the W4 method incorporates the dependence of the nonlinear source term on the solution more effectively.
The W4 iteration can be written in a general form as
\begin{eqnarray}
 \vx^{(k+1)} &=& \vx^{(k)} + \omega X\vp^{(k)},\nonumber\\
 \vp^{(k+1)} &=& \left(1-2\omega\right)\vp^{(k)}
  -Y \left( \vb(\vx^{(k)}) - A\vx^{(k)} \right),
\end{eqnarray}
where $X$ and $Y$ are matrices that are ideally derived from the Jacobian matrix.
In this setup, we choose $X=I$ and $Y=\tilde{D}^{-1}$
where $I$ is the identity matrix and $\tilde{D}$ is the diagonal part of the Jacobian matrix of the equation \eqref{eq:av-b}. Therefore, by computing the Jacobian matrix of the nonlinear system, the W4 method effectively incorporates the variation of the solution vector during the iteration. 
This leads to a more efficient update direction compared to the Jacobi method.

\medskip
\subsection{Restriction and Prolongation Operators}
In this subsection, we define the restriction and prolongation operators used in our code.

\begin{figure}
    \centering
    \includegraphics[scale=0.35]{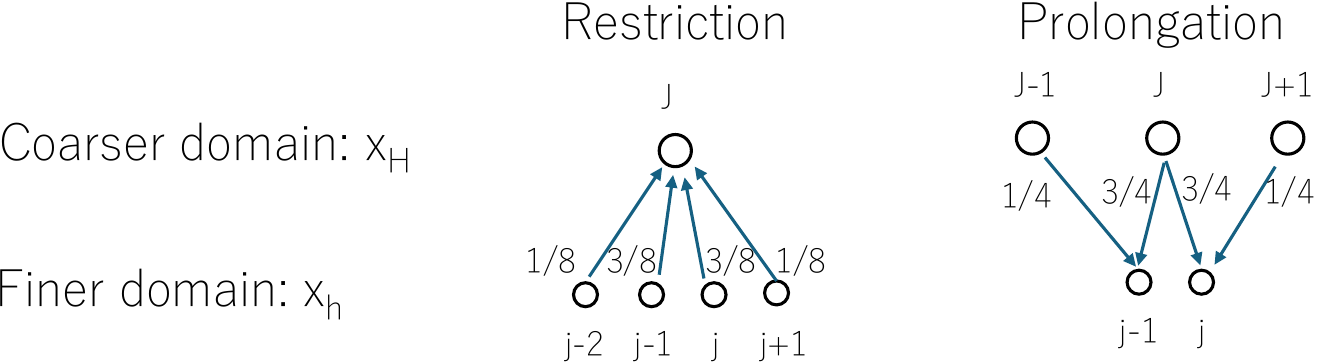}
    \caption{Schematic picture for restriction/prolongation in the multi-grid poisson solver.}
    \label{fig:MG}
\end{figure}

The restriction operator transfers information from the finer grid to the coarser grid~$x_H=I_h^Hx_h$.
Specifically, variables on the coarser grid are obtained using
a weighted averaging scheme by applying a tensor-product stencil
based on the one-dimensional weights:
\begin{eqnarray}
w_i\equiv\left(\frac{1}{8},\frac{3}{8},\frac{3}{8},\frac{1}{8}\right)
\end{eqnarray}
in each coordinate direction. For example, $x_H$ at the x-direction grid index $J$ is restricted by
\begin{align}
x_H|_J = \frac{1}{8} x_h|_{j-2} + \frac{3}{8}x_h|_{j-1} + + \frac{3}{8}x_h|_{j} + \frac{1}{8}x_h|_{j+1},
\end{align}
where the grid point $J$ in the coarser domain exists between the grid point $j-1$ and $j$ in the finer domain (see Fig.~\ref{fig:MG}).

In three dimensions, this results in a $4\times 4\times 4$ stencil
with weights:
\begin{eqnarray}
I_h^H = w_x w_y w_z.
\end{eqnarray}
This operator provides a high-order accurate averaging
from the finer grid to the coarse grid.
Alternatively, a lower-order restriction operator is defined as
a simple averaging over neighboring cells in the finer domain by applying a tensor-product stencil based on the one-dimensional weights:
\begin{eqnarray}
w_i\equiv \left(\frac{1}{2},\frac{1}{2}\right).
\end{eqnarray}
The use of a lower-order restriction helps avoid over-oscillation
and improves the solver's robustness.

The prolongation operator is constructed based on 
position-dependent interpolation on a cell-centered grid.
Each finer-grid point receives contributions from eight neighboring
coarse-grid points using direction-dependent weights.
For each coordinate direction, the interpolation weights
are given by
\begin{eqnarray}
v_i \in \left\{\frac{3}{4},\frac{1}{4}\right\},
\end{eqnarray}
depending on whether the fine-grid point is located closer to a coarse-grid node or not. For example, $x_h$ at the x-direction grid index $j-1$ and $j$ are prolonged, respectively, by
\begin{align}
x_h|_{j-1} &= \frac{1}{4}x_H|_{J-1} + \frac{3}{4}x_H|_J,\\
x_h|_{j} &= \frac{3}{4}x_H|_{J} + \frac{1}{4}x_H|_{J+1},
\end{align}
(see Fig.~\ref{fig:MG}).

The full three-dimensional interpolation obtained by a tensor-product
of the one-dimensional weights, 
resulting in a trilinear interpolation:
\begin{eqnarray}
I_H^h = v_x v_y v_z,~
x_h = I_H^h x_H.
\end{eqnarray}

This formulation is consistent with the cell-centered grid
arrangement, where finer and coarse grid points
do not coincide.

\medskip

The sample code is available in the public domain~\cite{Okawa2026}.
The CPU time to obtain the solution for the analytic problem in Appendix~\ref{appdx:NL} by the code is displayed
in Fig.~\ref{fig:MG_time}. 
\begin{figure}
    \centering
    \includegraphics[scale=0.35]{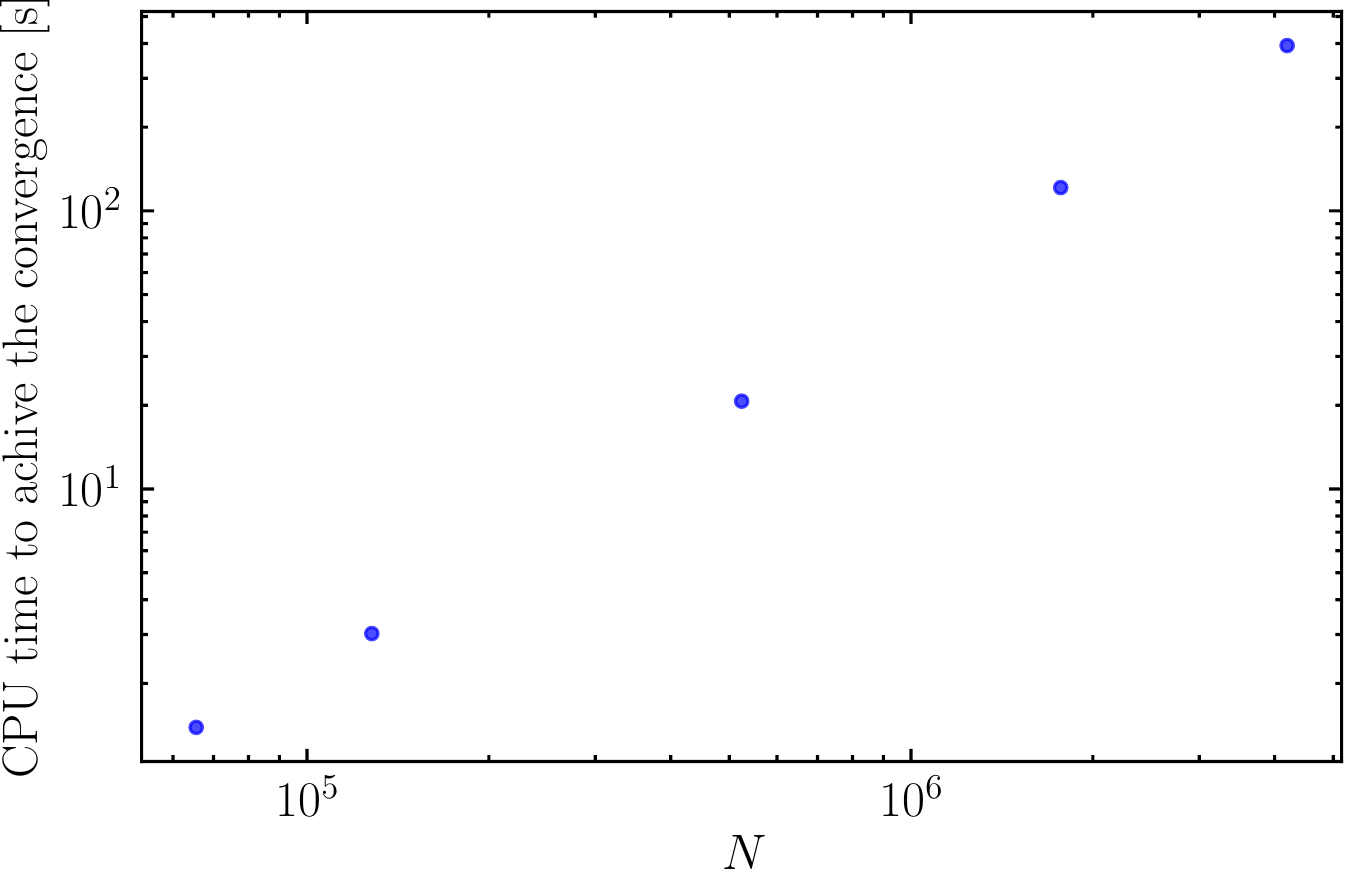}
    \caption{The required CPU time to achieve the convergence in the Newtonian constant density sphere as a function of the grid number}      \label{fig:MG_time}
\end{figure}

\section{Exact solution of the Newtonian constant-density sphere}\label{appdx:NL}
The exact solution of Eq.~(\ref{eq:Newtonian}) with $G=M=1$ unit is described as
\begin{align}
\phi(r) = \left\{\begin{array}{ll}
\frac{r^2}{2R_s^3} - \frac{3}{2 R_s} & (r \le R_s) \\
- \frac{1}{r} & (r > R_s).
\end{array}
\right.
\end{align}

\bibliography{reference}

\end{document}